\documentclass[prl,twocolumn,runinaddress,showpacs]{revtex4}
\usepackage{amsfonts}
\usepackage{amssymb}
\usepackage{amsmath}
\usepackage{graphicx}
\usepackage{bm}

\input{tcilatex}

\begin{document}

\title{$\mathbb{Z}_{2}$ invariant protected bound states in topological
insulators}
\date{\today }
\author{Wen-Yu Shan, Jie Lu, Hai-Zhou Lu and Shun-Qing Shen}
\affiliation{Department of Physics, The University of Hong Kong, Pokfulam Road, Hong Kong}

\begin{abstract}
We present an exact solution of a modifed Dirac equation for topological
insulator in the presence of a hole or vacancy to demonstrate that vacancies
may induce bound states in the band gap of topological insulators. They
arise due to the $\mathbb{Z}_{2}$ classification of time-reversal invariant
insulators, thus are also topologically-protected like the edge states in
the quantum spin Hall effect and the surface states in three-dimensional
topological insulators. Coexistence of the in-gap bound states and the edge
or surface states in topological insulators suggests that imperfections may
affect transport properties of topological insulators via additional bound
states near the system boundary.
\end{abstract}

\maketitle

Topological insulators are narrow-band semiconductors with band inversion
generated by strong spin-orbit coupling \cite{Moore-10Nature}. They are
distinguished from the ordinary band insulators according to the $\mathbb{Z}%
_{2}$ invariant classification of the gapped band insulators due to the time
reversal symmetry. The variation of the $\mathbb{Z}_{2}$ invariant at their
boundaries will lead to the topologically protected edge or surface states
with the gapless Dirac energy spectrum\cite%
{Kane-05prl,Fu06prb,Moore2007prbR,Roy2009prb,Fukui07jpsj,Qi08prb}.
Imperfections, such as impurity, vacancy, and disorder, are inevitably
present in topological insulators. Owing to the time-reversal symmetry, an
exciting feature of topological insulator is that its boundary states are
expected to be topologically protected against weak non-magnetic impurities
or disorders\cite{Wu06prl,Xu06prb}. This provoked much interest on the
single impurity problem on the surface of a topological insulator, starting
with gapless Dirac model\cite%
{Lee09prb,Zhou09prb,Guo10prb,Wang10prb,Biswas2010prb}. However, reminding
that the boundary state is only a manifestation of the topological nature of
bulk bands, it should also start with the examination of the host bulk to
know how the imperfections affect the electronic structure. It is well known
that single impurity or defect can induce bound states in many systems, such
as in the Yu-Shiba state in s-wave superconductor\cite{Yu-65,Shiba-68} and
in d-wave superconductors\cite{Balatsky-06rmp}. Topological defects were
discussed in the B-phase of $^{3}$He superfluid\cite{Volovik-03book} and
topological insulators and superconductors \cite{Teo-10prb}. Here we report
that bound states can form around a single vacancy in the bulk energy gap of
topological insulators. These bound states are found to have the same origin
as boundary states due to the $\mathbb{Z}_{2}$ classification, thus are also
topologically protected.

\begin{figure}[tbph]
\centering \includegraphics[width=0.5\textwidth]{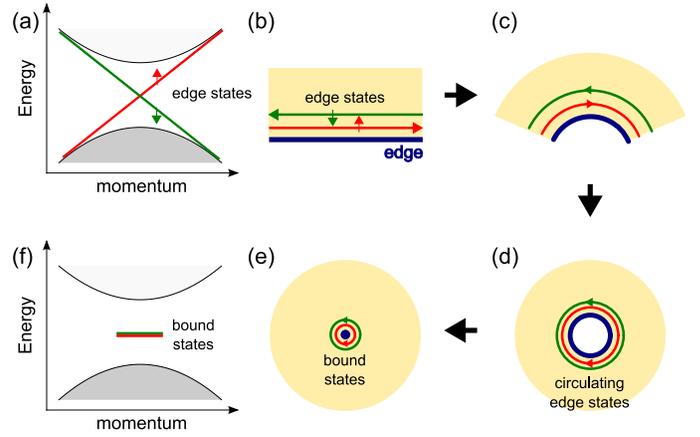}
\caption{Schematic description of the formation of vacancy-induced in-gap
bound states in two-dimensional topological insulators. [(a) and (b)] A pair
of helical edge states traveling along the edge of a 2D topological
insulator with the gapless Dirac dispersion. [(c) and (d)] When the edge is
bent into a hole, the helical edge states evolve to circulate around the
hole. [(e) and (f)] The circulating edge states may develop into
topologically-protected bound states as the hole shrinks into a point or
being replaced by a vacancy. The same physics is expected to happen in one
and three dimensions.}
\label{fig:schematic}
\end{figure}

The formation of the topological bound states can be readily illustrated by
reviewing the quantum spin Hall effect in two-dimensional (2D) topological
insulators\cite{Kane2005QSH,Bernevig-06Science,Konig2007science}, in which
strong spin-orbit coupling twists the bulk conduction and valence bands,
leading to a nontrivial $\mathbb{Z}_{2}$ index. As the $\mathbb{Z}_{2}$
varies across the edge, edge states arise in the gap with the gapless Dirac
dispersion. Unlike the quantum Hall effect in a magnetic field, spin-orbit
coupling preserves the time reversal symmetry, so the resulting edge states
appear in helical pairs, in which one state is the time-reversal counterpart
of the other, propagating along opposite directions and with opposite spins
(Fig. \ref{fig:schematic}). Now imagine that the system edge is rolled into
a hole, the edge states will circulate around the hole as the periodic
boundary condition along the propagating direction remains unchanged. While
shrinking its radius of the hole, most of the edge states will be expelled
into the bulk bands as the energy separation of the states becomes larger
and larger, and it is found that at least two degenerate pairs of the states
will be trapped to form the bound states in the gap while the hole will
evolve into a point defect. This mechanism of the formation of the bound
states can be realized in topological insulator in all the dimensions.

We will employ a modified Dirac model to provide a unified description of
topological insulators in various dimensions 
\begin{equation}
H_{0}=v\mathbf{p}\cdot \boldsymbol{\alpha }+\left( mv^{2}-Bp^{2}\right)
\beta .  \label{modified-Dirac}
\end{equation}%
The modification comes from the quadratic correction in momentum $-Bp^{2}$
to the band gap $mv^{2}$. $p_{i}=-i\hbar \partial _{i}$ is the momentum
operator ($i\in \{x,y,z\}$), $p^{2}=p_{x}^{2}+p_{y}^{2}+p_{z}^{2}$, $v$ and $%
m$ have the dimension of the speed and mass, respectively. $B$ has the
dimension of $m^{-1}$. The Dirac matrices satisfy the anticommutation
relations $\alpha _{i}\alpha _{j}=-\alpha _{j}\alpha _{i}$ ($i\neq j$)$,$ $%
a_{i}\beta =-\beta \alpha _{i}$ and $\alpha _{i}^{2}=\beta ^{2}=1$. One
representation of the Dirac matrices in three spatial dimensions can be
expressed as a set of $4\times 4$ matrices%
\begin{equation}
\alpha _{i}=\sigma _{x}\otimes \sigma _{i},\ \ \beta =\sigma _{z}\otimes
\sigma _{0},  \label{Dirac-matrices}
\end{equation}%
where $\sigma _{i=x,y,z}$ are the Pauli matrices, $\sigma _{0}$ is the $%
2\times 2$ unit matrix, and $\otimes $ represents the Kronecker product.
This modified Dirac Hamiltonian preserves the time reversal symmetry $\hat{%
\Theta}H_{0}\hat{\Theta}^{-1}=H_{0}$ under the time reversal operation $\hat{%
\Theta}=-i\alpha _{x}\alpha _{z}\hat{K}$, where $\hat{K}$ is the complex
conjugate operator. This equation has the identical mathematical structure
as the effective models for the quantum spin Hall effect and 3D topological
insulator\cite{Bernevig-06Science,Zhang2009NatPhys,Lu10prb,Shan10njp}. It
becomes topologically non-trivial if $mB>0$, while topologically trivial if $%
mB<0$ according to the $\mathbb{Z}_{2}$ classification of insulators\cite%
{Volovik-10xxx,Shen-10xxx}. Due to the bulk-boundary correspondence \cite%
{Qi06prb,Fu07prl}, there always exist topologically protected boundary
states at the open boundaries of a topological insulator, where the $\mathbb{%
Z}_{2}$ invariant changes from nontrivial to trivial. This feature can be
well described by the modified Dirac model when $mB>0$. Starting from this
modified Dirac model, we are now ready to explore existence of the in-gap
bound states induced by a single vacancy by presenting an exact solution of
the modified Dirac model.

\begin{figure}[tbph]
\centering \includegraphics[width=0.5\textwidth]{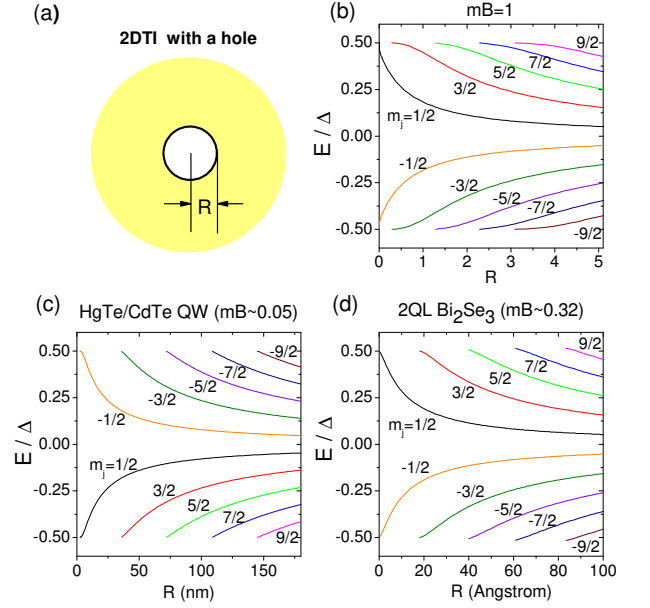}
\caption{Two-dimensional topologically protected bound states. (a) A 2D
topological insulator with a hole of radius $R$ at the center. (b)-(c),
Energies ($E$ in units of the band gap $\Delta $) of in-gap bound states
circulating around the hole as functions of the hole radius. $m_{j}$ is the
quantum number for z-component of total angular momentum of the circulating
bound states. In (b), $m=v=B=\hbar =1$ and $\Delta =\protect\sqrt{3}$ in
(c), $mv^{2}=-10$ meV, $B\hbar ^{2}=-686$ meV$\cdot $nm$^{2}$, and $\hbar
v=364.5$ meV$\cdot $nm, $\Delta =20$meV, adopted from Ref. \protect\cite%
{Bernevig-06Science}; in (d), $mv^{2}=0.126$ eV, $B\hbar ^{2}=21.8$ eV\r{A}$%
^{2}$, $\hbar v=0.294$ eV\r{A}, $\Delta =0.252$eV, adopted from Ref. 
\protect\cite{Zhang2010film}.}
\label{fig:2D}
\end{figure}

In two dimension ($p_{z}=0$) the equation can be reduced into two
independent $2\times 2$ hamiltonians 
\begin{equation}
h_{\pm }=(mv^{2}-Bp^{2})\sigma _{z}+\hbar v(p_{x}\sigma _{x}\pm p_{y}\sigma
_{y}),  \label{2D_Dirac}
\end{equation}%
with $h_{-}$ the time-reversal counterpart of $h_{+}$, which was already
studied in the HgTe quantum well\cite{Bernevig-06Science}, and in the thin
films of Bi$_{2}$Se$_{3}$\cite{Lu10prb,Shan10njp}. It is convenient to adopt
polar coordinates $(x,y)=r(\cos\varphi,\sin\varphi)$ in two dimensions. Here
these equations are solved under the vacancy boundary conditions [Fig. \ref%
{fig:2D}(a)], i.e., the center of the 2D topological insulator is punched
with a hole of radius $R$, thus the wavefunction is required to vanish at $%
r=R$ and $r=+\infty $. Due to the rotational symmetry, the z-component of
the total angular momentum $j_{z}=-i\hbar \partial _{\theta }+(\hbar
/2)\sigma _{z}$ provides a good quantum number, labeled by a half-integer $%
m_{j}\in \{\pm 1/2,\pm 3/2,...\}$, which can be used to characterize the
bound states. In this way the equation is reduced to a set of 1D radial
equations, which can be solved exactly. The trial wave function has the form 
$(\psi _{1},\psi _{2})^{\mathrm{T}}e^{\lambda r}$. The secular equations of
the indeterminate coefficients $(\psi _{1},\psi _{2})^{\mathrm{T}}$ give
four roots of $\lambda _{n}$ ($=\pm \lambda _{1},\pm \lambda _{2}$) in terms
of the energy $E$, 
\begin{equation}
\lambda _{1,2}^{2}=\frac{v^{2}}{2B^{2}\hbar ^{2}}[1-2mB\pm \sqrt{%
1-4mB+4B^{2}E^{2}/v^{4}}].  \label{lamda}
\end{equation}%
Using the boundary conditions at $r=R$ and $r=+\infty $ we finally arrive at
the transcendental equation for the bound state energies 
\begin{equation}
\frac{\lambda _{1}^{2}+\frac{mv^{2}-E}{B\hbar ^{2}}}{\lambda _{1}}\frac{%
K_{m_{j}+\frac{1}{2}}(\lambda _{1}R)}{K_{m_{j}-\frac{1}{2}}(\lambda _{1}R)}=%
\frac{\lambda _{2}^{2}+\frac{mv^{2}-E}{B\hbar ^{2}}}{\lambda _{2}}\frac{%
K_{m_{j}+\frac{1}{2}}(\lambda _{2}R)}{K_{m_{j}-\frac{1}{2}}(\lambda _{2}R)},
\label{2D_transcendental}
\end{equation}%
and the wavefunction $\Psi _{m_{j}}(r,\theta )$ for $h_{+}$ turns out to
have the form 
\begin{equation}
\left[ 
\begin{array}{c}
\frac{K_{m_{j}-\frac{1}{2}}(\lambda _{1}R)}{K_{m_{j}+\frac{1}{2}}(\lambda
_{1}R)}\left[ \frac{K_{m_{j}-\frac{1}{2}}(\lambda _{1}r)}{K_{m_{j}-\frac{1}{2%
}}(\lambda _{1}R)}-\frac{K_{m_{j}-\frac{1}{2}}(\lambda _{2}r)}{K_{m_{j}-%
\frac{1}{2}}(\lambda _{2}R)}\right] e^{i(m_{j}-\frac{1}{2})\theta } \\ 
i\frac{\lambda _{1}^{2}+\frac{mv^{2}-E}{B\hbar ^{2}}}{(\lambda _{1}v/B\hbar )%
}\left[ \frac{K_{m_{j}+\frac{1}{2}}(\lambda _{1}r)}{K_{m_{j}+\frac{1}{2}%
}(\lambda _{1}R)}-\frac{K_{m_{j}+\frac{1}{2}}(\lambda _{2}r)}{K_{m_{j}+\frac{%
1}{2}}(\lambda _{2}R)}\right] e^{i(m_{j}+\frac{1}{2})\theta }%
\end{array}%
\right] ,  \label{2d-solution}
\end{equation}%
where $K_{n}(x)$ is the modified Bessel function of second kind.

In Fig. \ref{fig:2D} (b)-(d), we show the bound-state energies as functions
of $R$ for an ideal case [(b), $mB=1$], for the HgTe quantum well ($mB=0.05$)%
\cite{Bernevig-06Science}, and for a 2 quintuple layer thick Bi$_{2}$Se$_{3}$
thin film ($mB=0.32$)\cite{Zhang2010film}. For a macroscopically large $R$,
we found an approximated solution for the energy spectrum of $h_{+}$ as $%
E=m_{j}\hbar v\mathrm{sgn}(B)/R$. As the time-reversal copy of $h_{+}$, $%
h_{-}$ has an approximated spectrum $E=-m_{j}\hbar v\mathrm{sgn}(B)/R$. They
form a series of paired helical edge states, in good agreement with the
edge-state solutions in a 2D quantum spin Hall system \cite{Zhou-08prl} if
we take $k=m_{j}/R$ for a large $R$. When shrinking $R$, the energy
separation of these edge state $\Delta E=\pm \hbar v\ /R$ increases with
decreasing $R$, and the edge states with higher $m_{j}$ will be pushed out
of the energy gap gradually. However, we observe that for $mB>0$ two pairs
of states for $m_{j}=\pm 1/2$ always stay in the energy gap, and as $%
R\rightarrow 0$ their energies approach to $E=\pm (v^{2}/2|B|)\sqrt{4mB-1}$
for $mB>1/2$ or $\pm mv^{2}$ for $0<mB<1/2$. The solutions demonstrate the
formation of the in-gap bound states as illustrated in Fig. \ref%
{fig:schematic}. Therefore considering the symmetry between $h_{+}$ and $%
h_{-}$ we conclude that there always exist \textit{at least} two degenerated
pairs of bound states in the energy gap in 2D quantum spin Hall system in
the presence of vacancy.

The mechanism of the formation of the in-gap bound states is applicable to
3D topological insulators.\cite{Xia2009.natphys.5.398,Zhang2009NatPhys} In
3D, the modified Dirac equation with a central potential becomes a classical
problem, the hydrogen atom-like problem. For the Coulomb potential, it was
exactly solved to give the fine structure of light spectra of hydrogen atom.
Similarly, the eigenstates of the 3D modified Dirac equation with a central
potential can be labeled by three good quantum numbers. The first two are
the total angular momentum $\mathbf{\hat{J}}=\mathbf{\hat{r}}\times \mathbf{%
\hat{p}}+\frac{\hbar }{2}\hat{\boldsymbol{\Sigma }}$ and its $z$-component $%
\hat{J}_{z}$, where the spin operator $\hat{\boldsymbol{\Sigma }}_{\alpha
}=\sigma _{0}\otimes \sigma _{\alpha }$ ($\alpha =x,y,z$). The eigenvalues
of $\mathbf{\hat{J}}^{2}$ and $\hat{J}_{z}$ are $j(j+1)\hbar ^{2}$ and $%
m_{j}\hbar $, respectively, with $j\in \{\frac{1}{2},\frac{3}{2},\cdots \}$
and $m_{j}\in \{-j,\cdots ,j\}$. The third conserved quantity is the
spin-orbit operator $\hat{\kappa}=\beta (\mathbf{\hat{r}}\times \mathbf{\hat{%
p}}\cdot \hat{\Sigma}+\hbar )$. Note that $\hat{\kappa}^{2}=\mathbf{\hat{J}}%
^{2}+\hbar ^{2}/4$, then the eigenvalues of $\hat{\kappa}$ is $\hbar \kappa
=\pm \hbar (j+1/2)=\pm \hbar ,$ $\pm 2\hbar ,\cdots $. Thus $\kappa $ here
is similar to the $\pm $ index in 2D that separate the hamiltonian into $%
h_{\pm }$. These conserved quantities also help to reduce the problem into a
set of 1D radial equations.\cite{Bjorken1964} In the presence of the vacancy
or a cavity of radius $R$ with the boundary conditions at $\Psi (R)=\Psi
(\infty )=0$, the radial part of the wave function can be solved in terms of
the modified spherical Bessel function of the second kind $k_{n}(x)$. With
the help of the recursion relation of $k_{n}(x)$, the transcendental
equations for the bound state energies can be found as 
\begin{equation}
\frac{\lambda _{1}^{2}+\frac{mv^{2}-E}{B\hbar ^{2}}}{\lambda _{1}}\frac{%
k_{j\pm \frac{1}{2}}(\lambda _{1}R)}{k_{j\mp \frac{1}{2}}(\lambda _{1}R)}=%
\frac{\lambda _{2}^{2}+\frac{mv^{2}-E}{B\hbar ^{2}}}{\lambda _{2}}\frac{%
k_{j\pm \frac{1}{2}}(\lambda _{2}R)}{k_{j\mp \frac{1}{2}}(\lambda _{2}R)}
\end{equation}%
for $\kappa =j+\frac{1}{2}$ and $-(j+\frac{1}{2})$, respectively. The
corresponding wavefunction $\Psi _{j,\kappa }^{m_{j}}(r,\theta ,\phi )$ are
of the form 
\begin{equation}
\Psi _{j,\kappa }^{m_{j}}(r,\theta ,\phi )\propto \left[ 
\begin{array}{c}
\frac{i(\lambda _{1}v/B\hbar )}{\lambda _{1}^{2}+\frac{mv^{2}-E}{B\hbar ^{2}}%
}[\frac{k_{j\mp \frac{1}{2}}(\lambda _{1}r)}{k_{j\mp \frac{1}{2}}(\lambda
_{1}R)}-\frac{k_{j\mp \frac{1}{2}}(\lambda _{2}r)}{k_{j\mp \frac{1}{2}%
}(\lambda _{2}R)}]\phi _{j,m_{j}}^{A/B} \\ 
\frac{k_{j\pm \frac{1}{2}}(\lambda _{1}R)}{k_{j\mp \frac{1}{2}}(\lambda
_{1}R)}[\frac{k_{j\pm \frac{1}{2}}(\lambda _{1}r)}{k_{j\pm \frac{1}{2}%
}(\lambda _{1}R)}-\frac{k_{j\pm \frac{1}{2}}(\lambda _{2}r)}{k_{j\pm \frac{1%
}{2}}(\lambda _{2}R)}]\phi _{j,m_{j}}^{B/A}%
\end{array}%
\right]
\end{equation}%
where 
\begin{eqnarray}
\phi _{j,m_{j}}^{A}(\theta ,\varphi ) &=&\left[ 
\begin{array}{c}
\sqrt{\frac{j+m_{j}}{2j}}Y_{j-\frac{1}{2}}^{m_{j}-\frac{1}{2}}(\theta
,\varphi ) \\ 
\sqrt{\frac{j-m_{j}}{2j}}Y_{j-\frac{1}{2}}^{m_{j}+\frac{1}{2}}(\theta
,\varphi )%
\end{array}%
\right] , \\
\phi _{j,m_{j}}^{B}(\theta ,\phi ) &=&\left[ 
\begin{array}{c}
-\sqrt{\frac{j-m_{j}+1}{2(j+1)}}Y_{j+\frac{1}{2}}^{m_{j}-\frac{1}{2}}(\theta
,\varphi ) \\ 
\sqrt{\frac{j+m_{j}+1}{2(j+1)}}Y_{j+\frac{1}{2}}^{m_{j}+\frac{1}{2}}(\theta
,\varphi )%
\end{array}%
\right] ,
\end{eqnarray}%
and $Y_{j}^{m}(\theta ,\varphi )$ is the spherical harmonics. $\phi
_{j,m_{j}}^{A}$ and $\phi _{j,m_{j}}^{B}$ possess opposite parities.

Although the rotational symmetry simplifies the problem, it is believed that
the presence of the bound states is not sensitive to the shape of the
vacancy, because of their topological origin. As an example we choose a set
of parameters based on first principles calculations for Bi$_{2}$Se$_{3}$ by
ignoring the anisotropy, where $mv^{2}=0.28eV$, $\hbar v=3.2$eV\AA , and $%
B=33$eV\AA $^{2}$. In this case $mB\sim 1>1/2$. Similar to the 2D case, we
find that the surface states around the cavity exist for a large radius $R$
as expected by the bulk-boundary correspondence for a $\mathbb{Z}_{2}$
invariant topological insulators\cite{Fu07prl}. The states with larger
orbital angular momentum are eventually expelled into the bulk band while
the radius is shrinking. We plot several bound state energies of small
orbital angular momenta as a function of the radius $R$ in Fig. \ref%
{fig:3D_hole}. For convenience, the bound states are labeled by the quantum
number $\kappa $ for the spin-orbit operator. Each $\kappa $ corresponds to $%
(2j+1)$-fold degenerate states of different $m_{j}$. Note that when the
vacancy radius is only several angstroms, two degenerate pairs of
bound-state energies can survive. Detailed analysis of the solution
indicates that the spatial distribution of a bound state is comparable with
that of the edge or surface states (for a large $R$ in the present case),
which are determined by the model parameters and slightly depends on $R$.
From the evolution of the edge or surface states into the in-gap bound
states, we think their formation have the same topological origin. Thus
these in-gap bound states are also protected topologically as the edge or
surface states.

\begin{figure}[tbph]
\centering \includegraphics[width=0.4\textwidth]{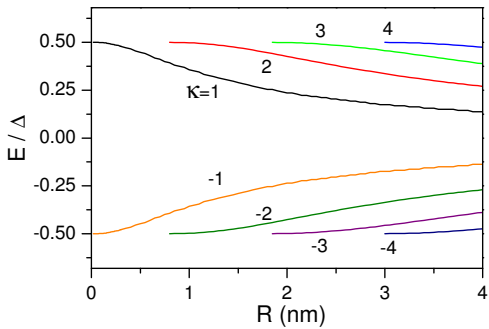}
\caption{Three-dimensional topologically protected bound states. Energies ($E
$ in units of the band gap $\Delta $) of in-gap bound states covering a
vacancy in a 3D topological insulator as functions of the vacancy radius $R$%
. $\protect\kappa $ is the quantum number of the spin-orbit operator.
Parameters: $mv^{2}=0.28$eV, $\hbar v=3.2$eV\r{A}, and $B\hbar ^{2}=33$eV\r{A%
}$^{2}$. $mB=0.90$ and $\Delta =0.50$eV. }
\label{fig:3D_hole}
\end{figure}

Now we come to address possible implication of these solutions to
topological insulators. Due to the overlapping in energy, when the vacancies
or defects are located close to the boundary, the induced in-gap bound
states may sabotage the electronic transport through the boundary states.
When the wave functions of the in-gap bound state and the edge or surface
states overlaps in the space, the distortion of the wave functions of these
states due to the boundary conditions will cause the energy change in these
states. As a result, we may also regard that there exists a transition
amplitude between these states. For example, in the 2D quantum spin Hall
system, there exists a pair of helical edge states with a linear dispersion, 
$E_{k,\pm }=\pm \hbar vk$. Consider a vacancy or defect appears near the
boundary. The effective model for a pair of helical edge states in the
presence of the defect states has the form, 
\begin{equation*}
H=\sum_{k,\sigma =\pm }\hbar vk\sigma c_{k\sigma }^{\dag }c_{k\sigma
}+\sum_{n}\epsilon _{n}d_{n}^{\dag }d_{n}+\sum_{k,\sigma ,n}(T_{k\sigma
}^{n}c_{k\sigma }^{\dag }d_{n}+h.c.)
\end{equation*}%
where $c_{k\sigma }^{\dag }$ and $c_{k\sigma }$ are the creation and
annihilation operators of edge states and $d_{n}^{\dag }$ and $d_{n}$ are
for the $n$th in-gap bound states with the energy $\epsilon _{n}$. The
resulting dispersions for the edge states are no longer linear in the
momentum, and open energy gaps $\Delta E=2\left\vert T_{k\sigma
}^{n}\right\vert $ at the resonant point $\epsilon _{n}=\hbar vk\sigma $. $%
T_{k\sigma }^{n}$ is a function of the relative position between the defect
or vacancy and the boundary. In the HgTe/CdTe quantum wells, a typical size
of the edge states and bound states is about 50 - 100 nm. The energy gap
opening here is \textit{not} caused by breaking the time reversal symmetry.
The mechanism is similar to the finite size effect in the quantum spin Hall
system, where the overlap of the edge states living on opposite edges causes
an energy gap\cite{Zhou-08prl}. In that case the energy splitting is about
0.5 meV for a strip system with a width of 200 nm. The effect is large
enough to be measured experimentally. This may help to understand why the
non-zero conductance is narrowed to a small region of gate voltage in the
HgTe/CdTe quantum wells\cite{Konig2007science}. The in-gap bound states may
also be one of the possible mechanisms for the low mobility in 3D
topological insulators.

However, blessings usually come in disguise. The whole semiconductor
business depends on how the positive and negative effects of impurities and
vacancies are precisely balanced. The topologically-protected bound states
for sure are essentially different from those we know before, as they are
subjected to some topological nature and confined to a mesoscopic scale.
Their possible impact and applications for topological insulators in future
deserve further studies to explore.

\textbf{Acknowledgements:} We would like to thank T. K. Ng for stimulating
our interest on this topic. This work was supported by the Research Grant
Council under Grant No. HKU 7051/10P and HKUST3/CRF/09.

\end{document}